\documentstyle[12pt]{article}
\begin{document}
\oddsidemargin 0.5cm
\evensidemargin 0.5cm
\topmargin -1cm
\baselineskip 16 true pt
\large
\title{Effect of memory and dynamical chaos in long Josephson junctions}
\author{K.N. Yugay, N.V. Blinov, and I.V. Shirokov \\
\em {Omsk State University,
Institute of Sensor Microelectronics}\\
\em{of Russian Academy of Sciences, Omsk 644077, Russia}}
\date{}
\maketitle
\begin{abstract}
A long Josephson junction in a constant external magnetic field and
in the presence of a dc bias current is investigated. It is shown that
the system, simulated by the sine--Gorgon equation, ``remembers'' a rapidly
damping initial perturbation and final asymptotic states are determined
exactly with this perturbation. Numerical solving of the  boundary
sine--Gordon problem and calculations of Lyapunov indices show
that this system has a memory even when it is in a state
of dynamical chaos, i.e., dynamical chaos does not destroy initial
information having a character of rapidly damping perturbation.\\
PACS number(s): 74.50+r, 05.45.+b
\end{abstract}

Dynamical chaos is one of the most interesting phenomena
in the theory of Josephson junctions.$^{1-10}$\
This phenomenon is not only
of theoretical importance but also of practical importance,
because many devices
are founded on Josephson junctions, in particular,
superconducting quantum interference devices (SQUID's).$^{11}$
Dynamical chaos in these  devices is another source of noise. Furthemore,
a long Josephson junction (LJJ) serves as a very good system for studying
nonlinear phenomena such as an exitation of fluxons and antifluxons,
their propagation, interaction, scattering, and breakup. Investigations
of the last few years showed that a LJJ detects deeper characteristics
that had seemed. Even in the simplest case, when a bias
current and an external oscillating field are absent, the presence only
of a constant external magnetic field leads to the most interesting phenomenon
connected with the selection of the solution of the stationary
Ferrell-Prange equation. The fact is that this equation has not only
provided one solution by given boundary conditions;
the number of these solutions
increases by the strength of the external magnetic field and the total length
of the junction.$^{11}$

Recently we have shown$^{12}$ that the selection of a solution is
carried out with the form of a small and rapidly damping initial
perturbation in time in the nonstationary sine-Gordon equation and what
is more surprising an asymptotic solution of this equation coincides
with one of
the stable solutions of the stationary Ferrell-Prange equation. Two
circumstances are remarkable here: (1) A small perturbation influences
very much the evolution of the system with $t\to\infty$;
in a sense it defines the character of
asymtotic solutions. (2) One can say that in spite of the fact that
a small perturbation is a rapidly damping one, the stable asymptotic
solution ``remembers'' the initial perturbation.
In other words, the nonlinear system, i.e., a LJJ, described with the
sine-Gordon equation shows an effect of memory. However,
in Ref. 12, the  LJJ is studied solely under the influence of an external
constant magnetic field. Therefore, it is of interest to investigate
the LJJ from the point of view of the effect of memory not only in the
presence of an external constant magnetic field
but also under the influence of
a dc bias current through the junction  causing an exitation of
dynamical chaos. How will the effect of memory in the presence
of a dc bias current be shown? Will this effect take place in the
states of dynamical chaos in general? Below we will try to
give answers on these questions.

We write down the sine-Gordon equation in the presence of a dc bias
current in a LJJ in the form
\begin{equation}
  \varphi_{tt}(x,t)+2\gamma\varphi_t (x,t) -
  \varphi_{xx}(x,t)=-\sin\varphi(x,t)+\beta,
\end{equation}
where $\varphi(x,t)$ is the Josephson phase variable,
$x$ is the distance along the junction normalized to the Josephson
penetration length $\lambda_J$,
$$
  \lambda_J=\Bigl({c\Phi_0\over 8\pi^2j_cd}\Bigr)^{1/2},
$$
$\Phi_0$ is the flux quantum,
$j_c$ is the critical current density of the Josephson junction,
$d=2\lambda_L+b$, $\lambda_L$ is the London penetration length, $b$ is the
thickness of the dielectric barrier,
$t$ is the time normalized to the inverse of the
Josephson plasma frequency $\omega_J$,
$$
  \omega_J=\Bigl({2\pi c j_c\over C\Phi_0}\Bigr)^{1/2},
$$
$C$ is the junction capacitance per unit area,
$\gamma$ is the dissipative coefficient per unit area, and $\beta$ is the
dc bias current density normalized to $j_c$.

We write down the boundary condition for Eq. (1) in the form
\begin{equation}
  \begin{array}{c}
     {\partial\varphi(x,t)\over\partial x}\vert_{x=0}\equiv
      H(0,t) =
     {\partial\varphi(x,t)\over\partial x}\vert_{x=L}
      \equiv H(L,t)\\=H_0\left(1-a e^{-t/t_0}\cos t\right),
   \end{array}
\end{equation}
where $L$ is the total length of the junction normalized to $\lambda_J$,
$H_0$ is the external constant magnetic field perpendicular to the
junction and normalized as well as $H(0,t)$ and $H(L,t)$ to
${\Phi_0\over 2\pi\lambda_J d}$, $a$ is the controlling
(perturbation) parameter characterizing the rapidly damping in the time
perturbation, and $t_0$ is the characteristic time of this damping
perturbation normalized to $\omega^{-1}_J$.

Eq. (1) with boundary condition (2)  is solved numerically.
In contrast to the case $\beta=0$ considered by us$^{12}$, the
picture of magnetic field evolution in the junction turns out
more complicated by $\beta\neq 0$ as will be shown below. Physically
this is connected with the fact that the energy balance in the junction
in the presence of bias current is such that the energy brought into the
system with this current can make up for the energy loss because
of dissipation or exceed it. And so one can expect that at few
values of $\beta$ a state of junction will differ little from the stationary
one described by the Ferrell-Prange equation. At sufficiently large
values of $\beta$ asymptotic regular (periodic) solutions that
represent waves --- fluxons and antifluxons --- moving along the junction
and interacting among themselves and with junction boundaries and also
nonregular solutions that represent dynamical chaos will take place$^1$.
Our calculations showed that if an asymtotic state at $a=0$ (further
we shall call the state at $a=0$ as ``starting'') is regular, by
introduction at the initial moment of a rapidly damping perturbation
defined with the controlling parameter $a$ as well as in a stationary
case, examined in Ref. 12, the selection of the asymtotic solution
by a given set of parameters $H_0$, $\gamma$, $L$, and $\beta$ is determined
with this parameter $a$; i.e., the system ``remembers'' the form of the
rapidly damping perturbation and chooses the way of further evolution in
accordance with this. (It is noteworthy that the
effect of memory discussed here happens in a dissipative system and so
it is not
connected with the reproduction of a signal as it takes place,
for example, in noncollision plasma in the effect of plasma echo$^{14}$).
Furthermore, our calculations show that if the ``starting'' state is
regular (periodic), stationary states can also arise by introduction
of perturbation ($a\neq 0$) which it is of surprise in itself.
However, the most remarkable fact is that the system is very sensitive
to a rapidly damping perturbation as well as in the dynamical chaos
conditions; i.e., the system has memory in this case too.

For the quantitative description of different characteristic states we
used Lyapunov indices. We write down Eq. (1) in the form
\begin{equation}
 \left\{\begin{array}{lll}\varphi_t&=&V,\\
           V_t&=&-2\gamma V+\varphi_{xx} -\sin\varphi+\beta,
        \end{array}
 \right.
\end{equation}
or, that is the same, in the form
\begin{equation}
z_t = F(z),
\end{equation}
where $z$ is the vector with the components $\varphi$ and $V$,
$z\equiv\left(\begin{array}{c}\varphi\\V\end{array}\right)$,
and $F(z)$ is defined as follows:
$$
  F(z)\equiv F(\varphi,V)=
  \left(\begin{array}{l}V\\
  -2\gamma+\varphi_{xx}-\sin\varphi+\beta\end{array}\right).
$$
Let $z(t)$ be a solution of Eq. (4). Then we can write down the
equation for variations:
\begin{equation}
   \begin{array}{lll}
       w_t&=&{\partial F\left(z(t)\right)\over\partial z}w,\\
       w&\equiv&\left(\begin{array}{c}w_1\\w_2\end{array}\right).
   \end{array}
\end{equation}
We define the Lyapunov index (LI) as
\begin{equation}
 \lambda = \lim_{t\to\infty}{1\over t}\ln{||w(t)||\over||w(0)||},
\end{equation}
where $||w||$ is the vector norm that we define as Euclidean norm
\begin{equation}
      ||w||^2=\int\limits^L_0(w^2_1+w^2_2)\,d x.
\end{equation}
Depending on direction of the initial vector $w(0)$, different LI's
will exist and their number will be infinite. The definition of
LI (6) for system (3) is a natural generalization of a LI
for finite-dimensional dynamic systems.$^{13}$

The maximum LI plays a very important role because it is precisely this
maximum that determines the motion character --- exponential growth, decay,
or zero change --- for the majority of the trajectories of the system. A set of
initial data $w(0)$ for which formula (6) gives LI's differing from
the maximum one is negligibly small and by numerical calculations this formula
gives the maximum LI as a rule.

The results a numerical solution of Eq. (1) with boundary conditions (2)
and calculation of LI (6) showed that there exist three forms of characteristic
states of the system for which the maximum LI can be as follows:
(1)$\lambda > 0$, (2)$\lambda < 0$, and (3)$\lambda\leq 0$. The states
with $\lambda > 0$ represent the dynamical chaos states (Fig. 1), the states
with $\lambda < 0$ represent the stable stationary states (Fig.2), and
the states with $\lambda\leq 0$ represent the regular (periodic)
states (Fig. 3). All states in Figs. 1--3 that are shown as illustration
are as ``starting'' ones ($a=0$) and they are calculated with identical
values of the parameters $H_0=1.25$, $L=5$, and $\gamma=0.26$, but
with different values of $\beta$. For the chaos state in Fig. 1, $\beta=0.50$;
for the stationary one in Fig. 2, $\beta=0.427$; and for the
regular one in Fig. 3, $\beta=0.60$. In Figs. 1--3 the dependences of
the potential $\varphi_t$ (potentials are normalized to the value
$V_c\equiv{\hbar\omega_p\over 2e}$) on time and
the calculated values LI $\lambda$ corresponding
to them are shown. We note that the
``starting'' chaotic state, represented in Fig. 1, is the same as in Ref. 1.

Let us examine the specific set of parameters, i.e., the definite point
in parameter space, corresponding to the chaotic ``starting'' state:
$H_0=1.25$, $L=5$, $\gamma=0.26$, $\beta=0.44$, and $a=0$. If we input now
the rapidly damping in the time perturbation determined by the controlling
parameter $a$, the system does not remain
in the previous chaotic state as the calculations show, but wanders
between all three forms of
states: chaotic, stationary, and regular, when this parameter changes.
By calculations the following hierarchy of times was used:
$t_0\ll \tau_r\ll T$, where $\tau_r$ is the characteristic time of
relaxation processes ($\tau_r$ is the time of relaxation to asymtotic
states) and $T$ is the time of observation. The values of characteristic
times in our calculations were as follows: $t_0=5$,
$\tau_r\approx 60$, $T=2000$. At first sight, one might have expected
that initial perturbations at the time interval $T$ damping at the time about
$t_0$ would be forgotten and they will have no influence on the evolution
in the large time interval. [We notice that $\tau_r$ is not equal to
$\gamma^{-1}$, exactly $\tau_r\gg \gamma^{-1}$ in our case.
This is connected with the system's nonlinearity. The value of $\tau_r$
is defined from numerical calculation of our problem (1), (2).]
In Fig. 4 are shown the results of calculation of LI's with values
of parameters for
$H_0$, $L$, $\gamma$, and $\beta$ mentioned above and by specific
values of parameter $a$. As we can see from Fig. 4 three typical clusters of
states take place: a cluster of chaotic states {\it ch}
(in Fig. 4 the following
values of the parameter $a$ correspond to them: $a$ =0, 0.175,
0.180, 0.280), a cluster of regular states {\it r} ($a$ = 0.290, 0.300,
0.320), and a cluster of stationary states {\it s} ($a$ = 0.190, 0.195,
0.285). For the cluster {\it ch}
the values of the LI $\lambda\approx 5\cdot 10^{-2}$;
for the cluster {\it r}, $\lambda\approx -10^{-3}$; and for the cluster
{\it s}, $\lambda\approx -10^{-1}$.

In Fig. 5 the potentials on the junction $\varphi_t$ depending
on time for three values of parameter $a$ differing from each other
on 0.005 and belonging to cluster {\it ch}, $a=0.280$ [Fig. 5a];
to cluster {\it s}, $a=0.285$ [Fig. 5b]; and to cluster {\it r},
$a=0.290$ (Fig. 5c)
are showed; their LI's are represented in Fig. 4. Thus, a small
change of parameter $a$ leads to a transition between all three
characteristic states of the system.

In Table I transitions between chaotic {\it ch}, stationary {\it s},
and regular {\it r}
states are represented by a change of the parameter $a$ from 4.000 to 4.155
at the fixed remaining parameters indicated above.
\begin{center}
Table I. States of a LJJ

\vspace{0.5 true cm}
\begin{tabular}{|c|c|c|c|c|c|c|c|}\hline
4.000   &4.005   &4.010   &4.015   &4.020   &4.025   &4.030   &4.035\\
\hline
{\it r} &{\it r} &{\it r} &{\it r} &{\it r} &{\it r} &{\it r} &{\it ch}\\
\hline
\end{tabular}

\vspace{0.5 true cm}
\begin{tabular}{|c|c|c|c|c|c|c|c|}\hline
4.040    &4.045    &4.050    &4.055    &4.060   &4.065   &4.070   &4.075\\
\hline
{\it ch} &{\it ch} &{\it ch} &{\it s}  &{\it s} &{\it s} &{\it s} &{\it s}\\
\hline
\end{tabular}

\vspace{0.5 true cm}
\begin{tabular}{|c|c|c|c|c|c|c|c|}\hline
4.080    &4.085    &4.090    &4.095    &4.100    &4.105    &4.110
&4.115\\ \hline
{\it ch} &{\it ch} &{\it ch} &{\it ch} &{\it ch} &{\it ch} &{\it ch}
&{\it ch}\\ \hline
\end{tabular}

\vspace{0.5 true cm}
\begin{tabular}{|c|c|c|c|c|c|c|c|}\hline
4.120   &4.125   &4.130   &4.135   &4.140   &4.145   &4.150   &4.155\\
\hline
{\it s} &{\it s} &{\it r} &{\it r} &{\it r} &{\it r} &{\it r} &{\it r}\\
\hline
\end{tabular}
\vspace{0.5 true cm}
\end{center}

We note that the transitions between the states, reduced in Table I
and stipulated by a change of the perturbation parameter $a$, correspond
to the ``starting'' state of chaos ($a=0$). Thus, by the given
values of $H_0$, $L$, $\gamma$, and $\beta$ a final asymptotic state is
determined with the parameter $a$ independently of this, whether
the ``starting'' state is chaotic or not. Such asymptotic behavior
of the system says that dynamical chaos essentially differed from statistical
chaos by which  any perturbation damps rapidly and a system relaxes to its
final state (for example, to the state of thermodynamical equilibrium)
completely ``having forgotten'' an initial perturbation; i.e., the final
state does not depend on this perturbation. As we see, a system in the state
of dynamical chaos in contrast to a system being in the state of
statistical chaos ``remembers''  the initial perturbation and, in any
sense, final states and transitions between them are defined with the
very initial perturbation. It makes it possible for us to recognize this memory
effect in the system described with the sine-Gordon equation with
dissipation and in the presence of an external magnetic field and
a bias current. Thus, the dynamical chaos originating in
a nonlinear system does not destroy an initial information;
i.e., the nonlinear system has a memory in the states of dynamical
chaos as well.

\vspace{1 cm}
\noindent
{}$^1$W.J. Yeh, O.G. Symko, and D.J. Zheng, Phys. Rev. {\bf B 42},
4080(1990).\\
{}$^2$L.E. Guerrero and M. Ostavio, Physica B {\bf 165-166}, 1657(1990).\\
{}$^3$L.E. Guerrero and M. Ostavio, Physica B {\bf 165-166}, 1659(1990).\\
{}$^4$M. Cirillo and N.F. Pedersen, Phys. Lett. A {\bf 90}, 150(1982).\\
{}$^5$N. Gronbech-Jensen, P.S. Lomdahl, and M.R. Samuelsen,
Phys. Rev. {\bf B 43}, 12799(1991).\\
{}$^6$N. Gronbech-Jensen, Phys. Rev. {\bf B 45}, 7315(1992).\\
{}$^7$S. Rajasekar and M. Lakshmanan, Physica A {\bf 167}, 793(1990).\\
{}$^8$S. Rajasekar and M. Lakshmanan, Phys. Lett. A {\bf 147}, 264(1990).\\
{}$^9$E.F. Eriksen and J.B. Hansen, Phys. Rev. {\bf B 41}, 4189(1990).\\
{}$^{10}$X. Yao, J.Z. Wu, and C.S. Ting, Phys. Rev. {\bf B 42}, 244(1990).\\
{}$^{11}$A. Barone and G.Paterno,{\em Physics and Applications
of the Josephson Effect} (Wiley-Interscience,New-York, 1982).\\
{}$^{12}$K.N. Yugay, N.V. Blinov, and I.V. Shirokov,
Phys. Rev. {\bf B 49}, 12036 (1994).\\
{}$^{13}$A.J. Lichtenberg and M.A. Lieberman, {\em Regular and Stochastic
Motion} (Springer-Verlag, New-York, 1983).\\
{}$^{14}$F.F. Chen, {\em Introduction to Plasma Physics and Controlled
Fusion} (Plenum Press, New-York, 1984).\\

\newpage\thispagestyle{empty}
\center{\bf Effect of memory and dynamical chaos\\ in long Josephson junctions}
\center{K.N. Yugay, N.V. Blinov, and I.V. Shirokov}
\vspace{1cm}
\begin{enumerate}
\item[Fig. 1.] The potential $\varphi_t$ (a) and the Lyapunov index (b)
               in a chaotic state by $\beta=0.50$. The values of other
               parameters are $H_0=1.25$, $L=5$, $\gamma=0.26$, $a=0$.
\item[Fig. 2.] The potential $\varphi_t$ (a) and the Lyapunov index (b)
               in a stationary state by $\beta=0.427$. Other
               parameters are the same as those in Fig. 1.
\item[Fig. 3.] The potential $\varphi_t$ (a) and the Lyapunov index (b)
               in a regular state by $\beta=0.60$. Other
               parameters are the same as those in Fig. 1.
\item[Fig. 4.] The Lyapunov indices $\lambda$ vs parameter $a$:
               {\it ch} is the cluster of chaotic states ($a=$0, 0.175,
               0.180, 0.280), {\it r} is the cluster of regular states
               ($a=$0.290, 0.300, 0.320), and {\it s} is the cluster of
               stationary states ($a=$0.190, 0.195, 0.285).
               The values of other parameters are just the
               same: $H_0=1.25$, $L=5$, $\gamma=0.26$, $\beta=0.44$.
\item[Fig. 5.] The potential $\varphi_t$ vs $t$ belonging to the cluster
               {\it ch}, $a=0.280$ (a), to the cluster {\it s}, $a=0.285$ (b),
               and to the cluster {\it r}, $a=0.290$ (c). Other
               parameters are the same as those in Fig. 4.
\end{enumerate}
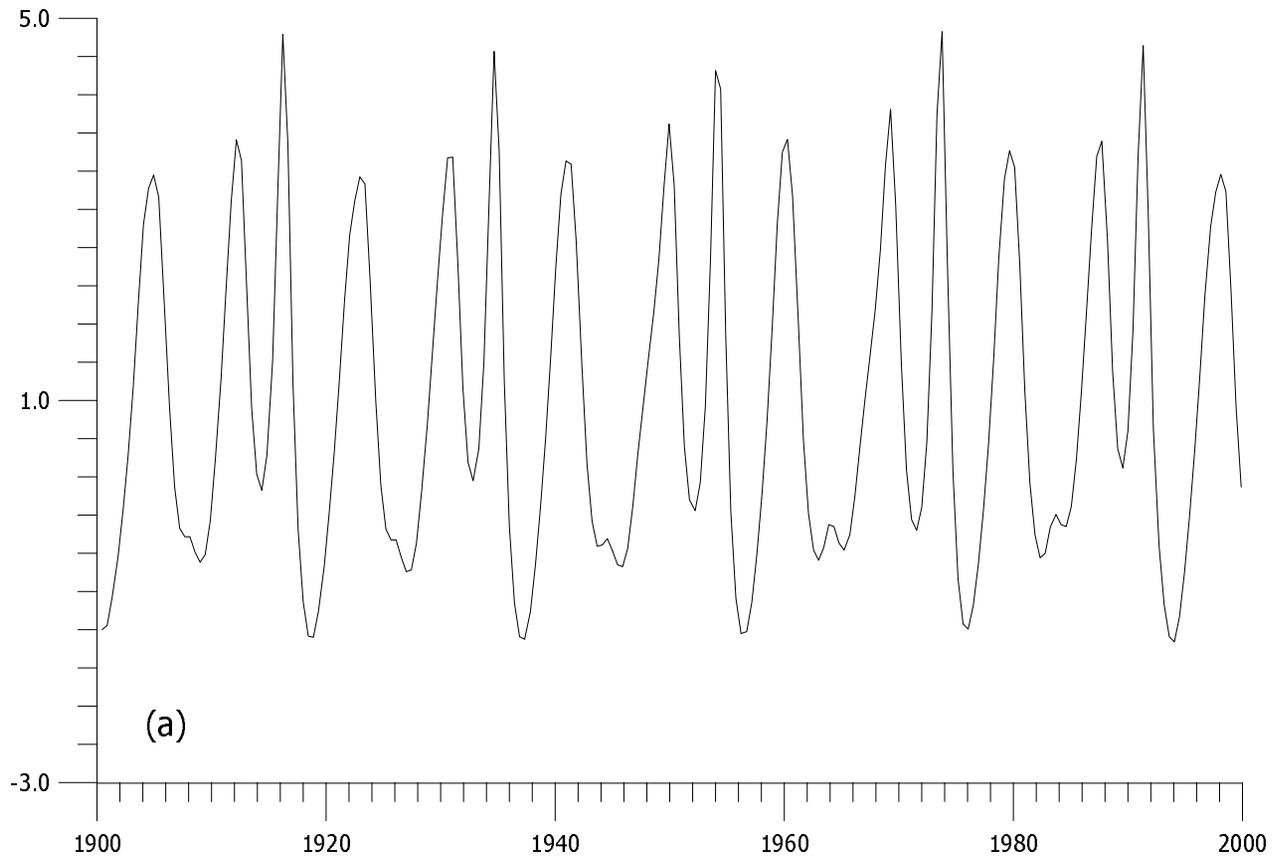
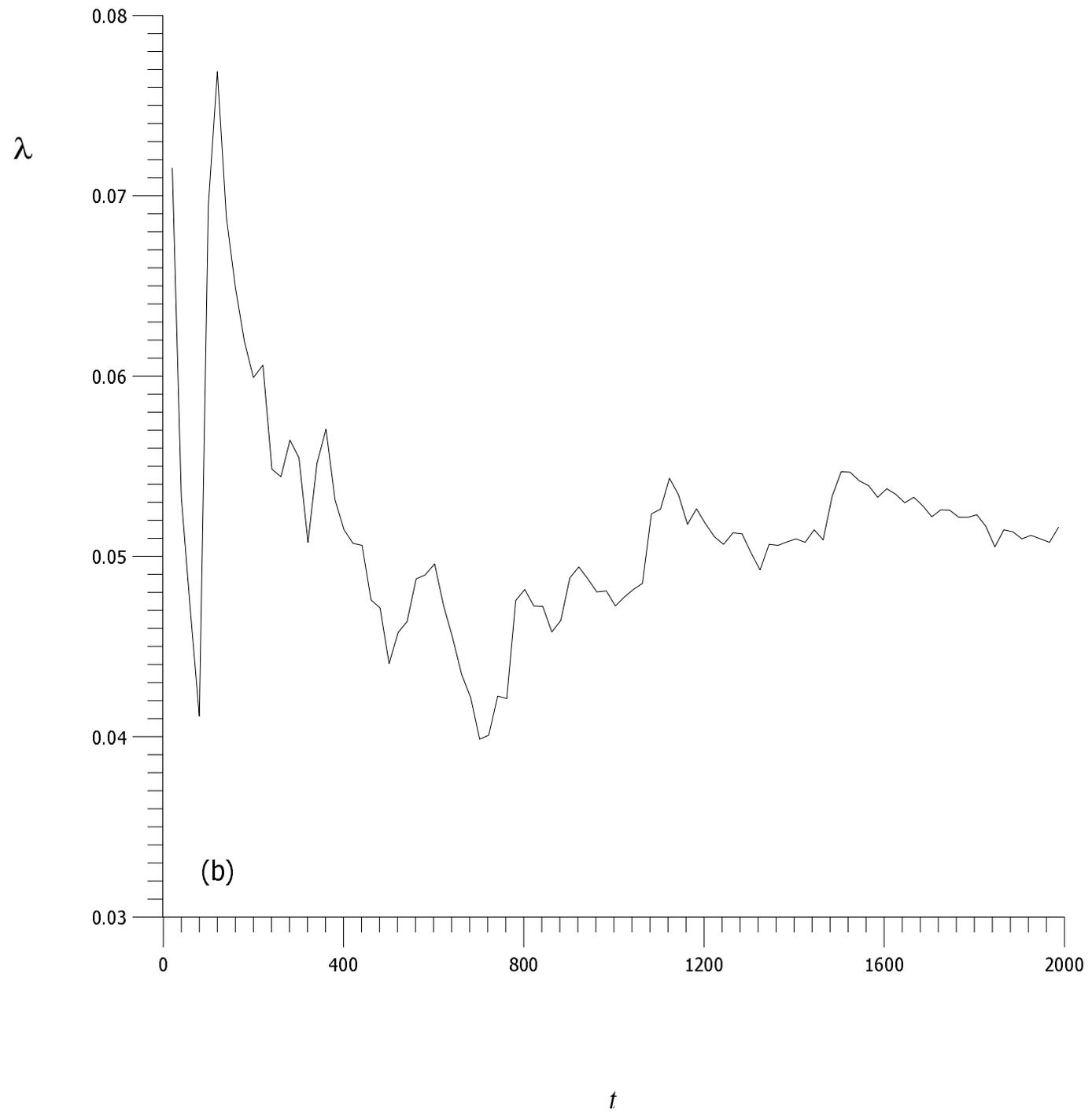
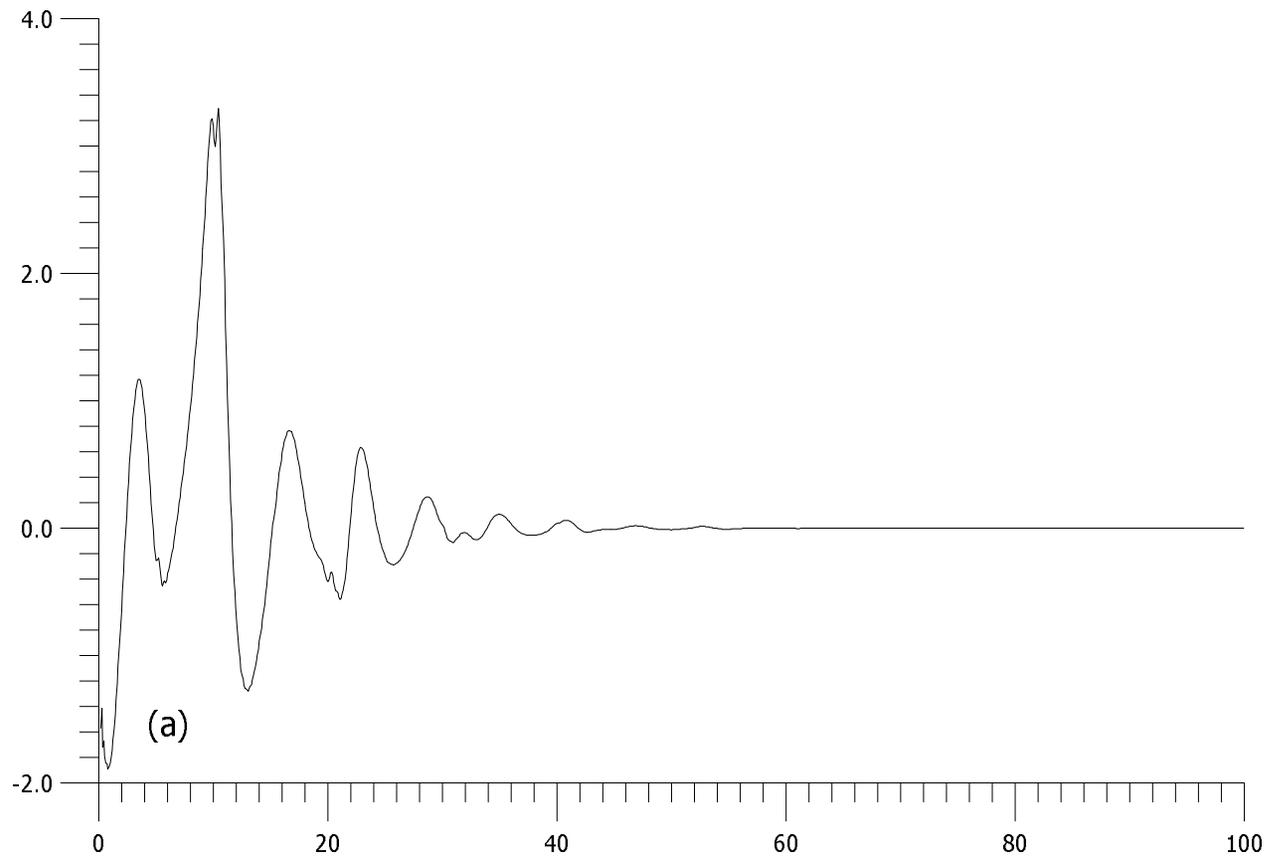
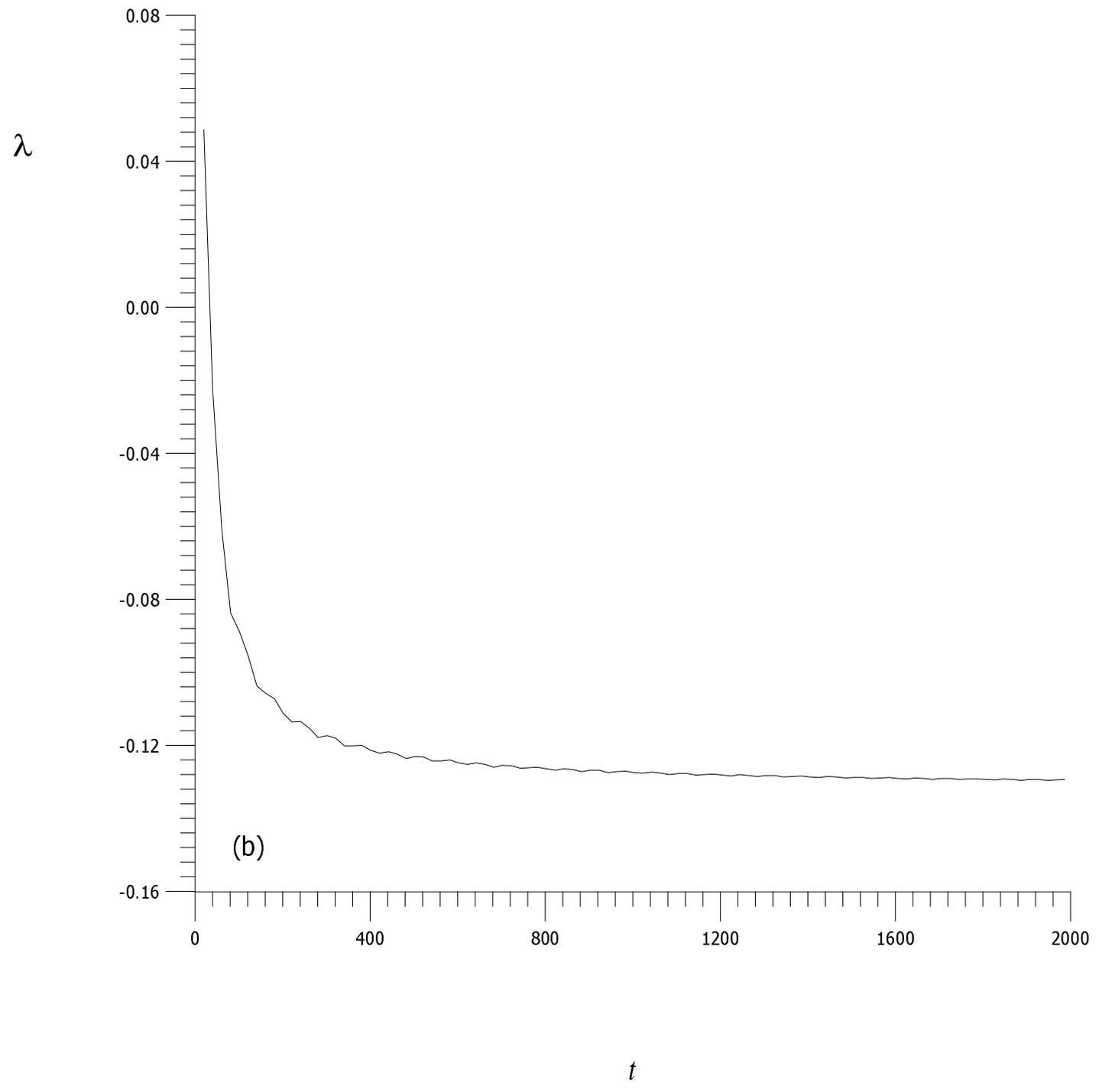
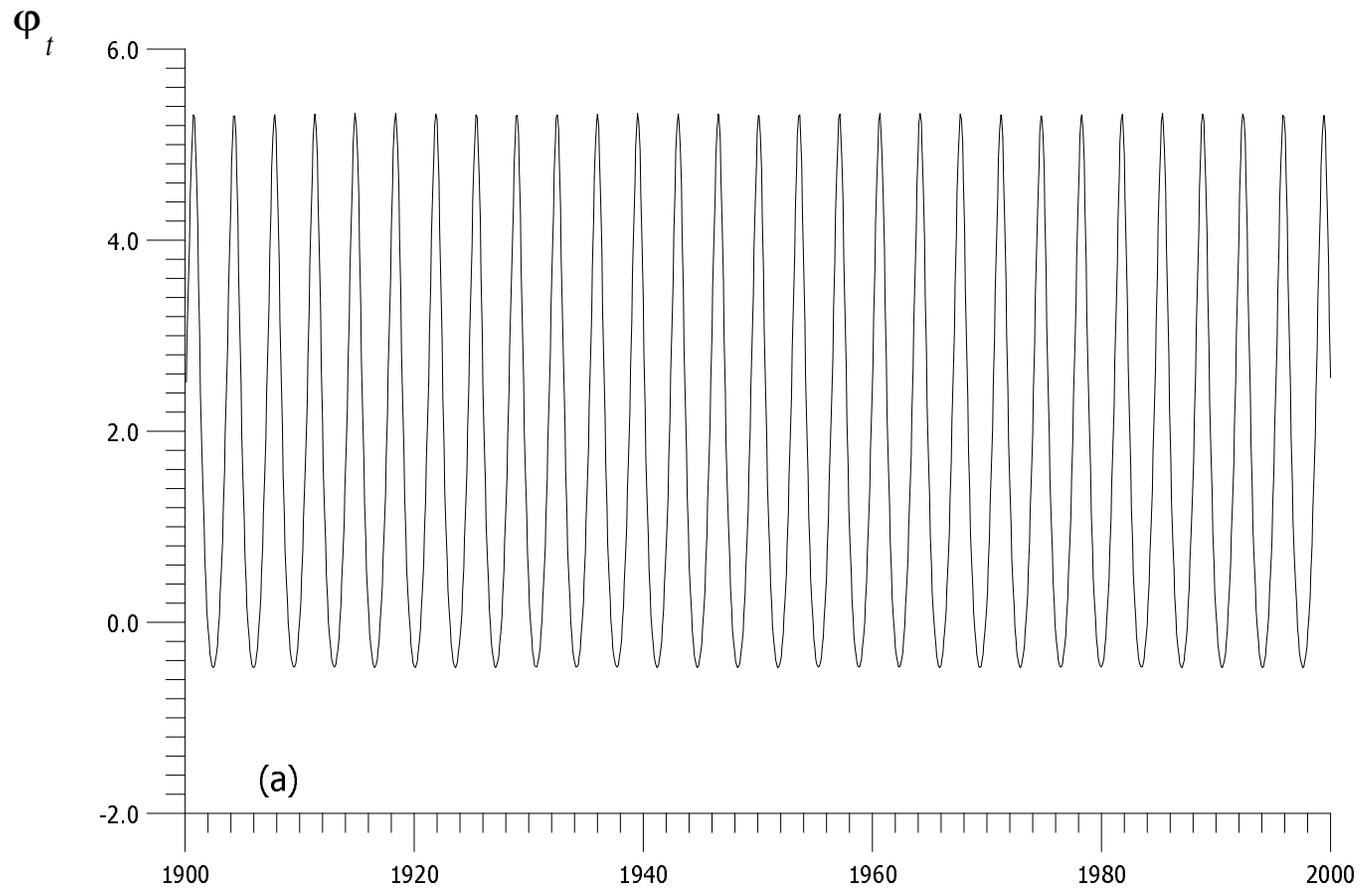
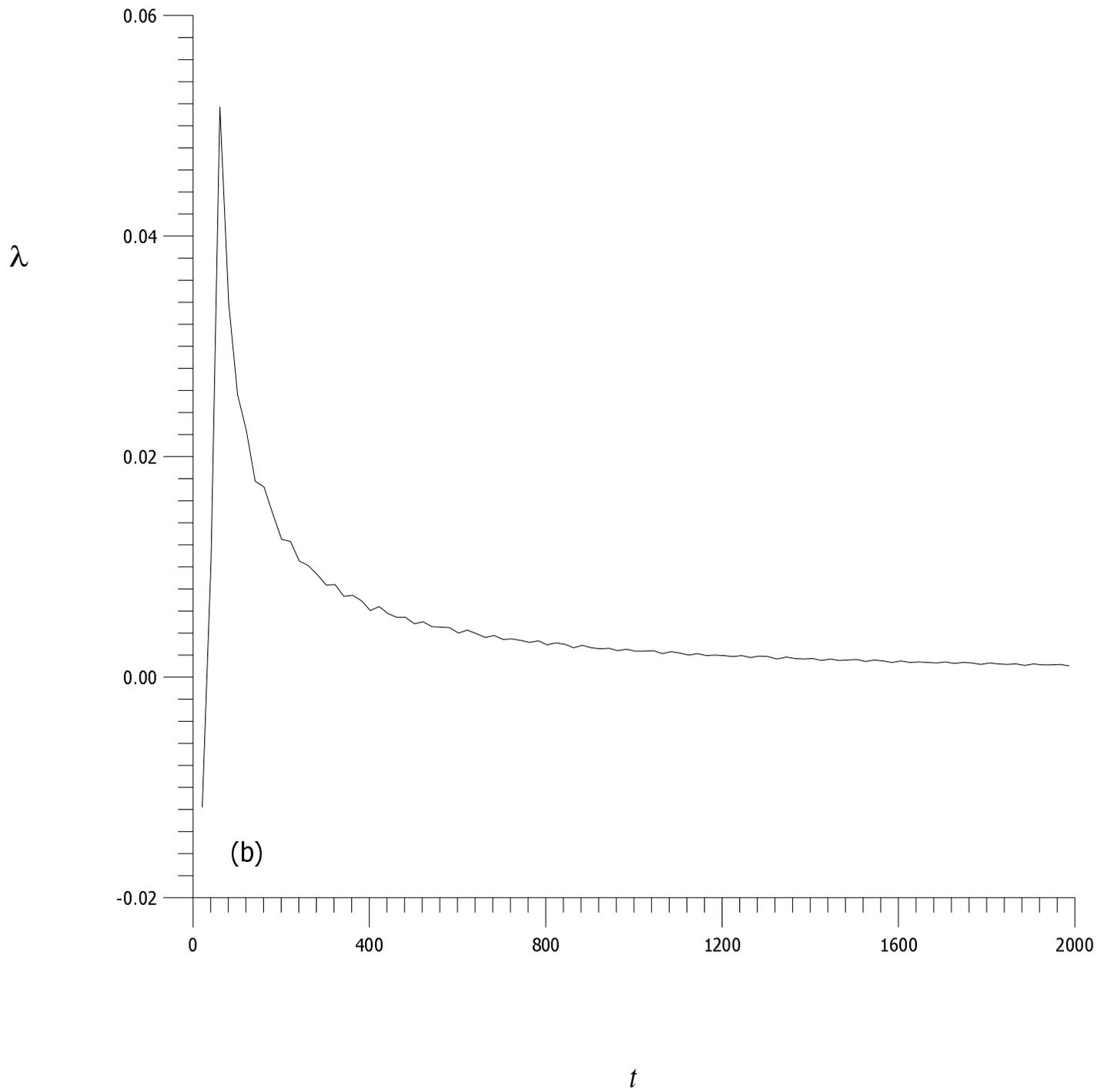
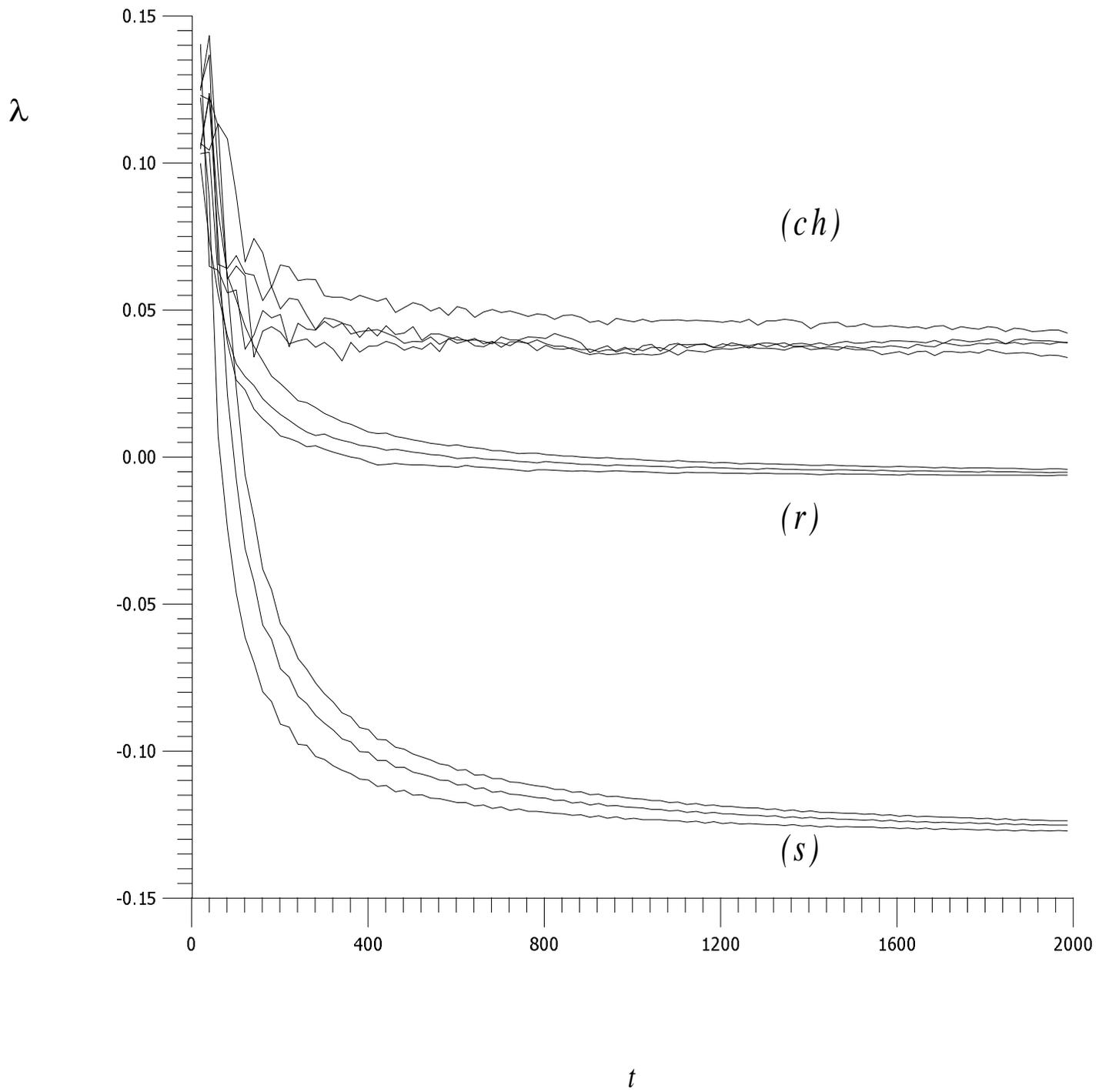
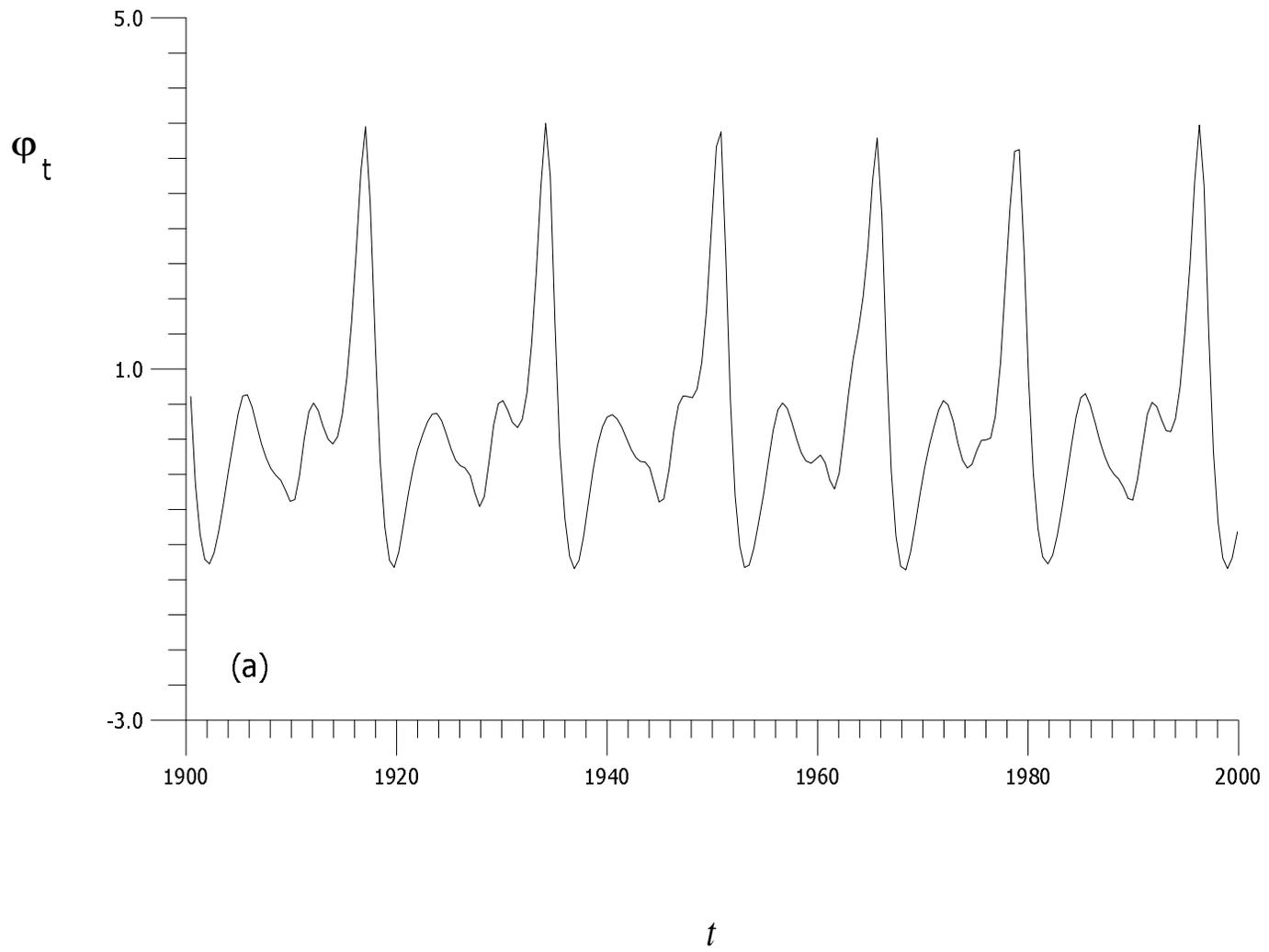
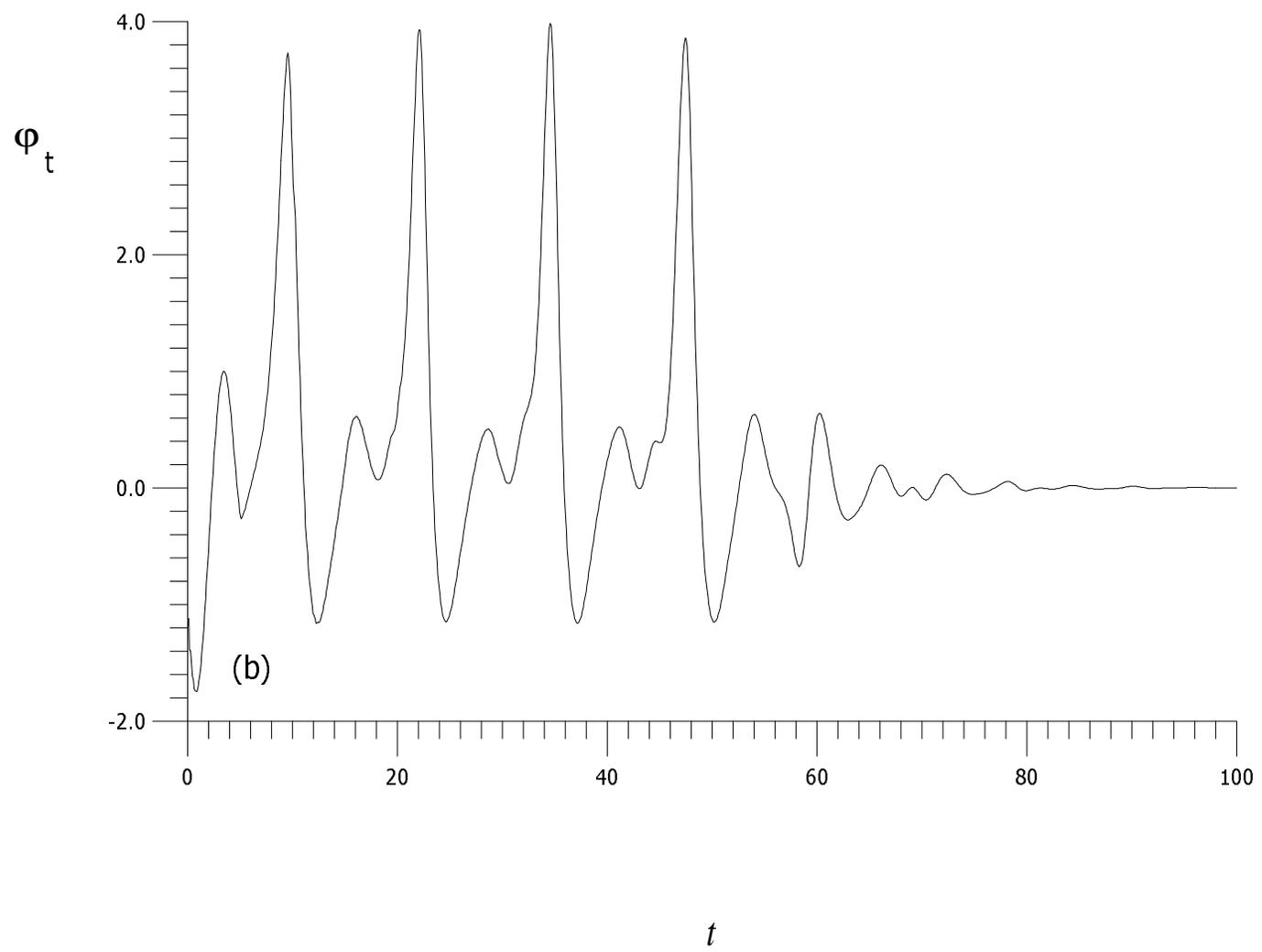
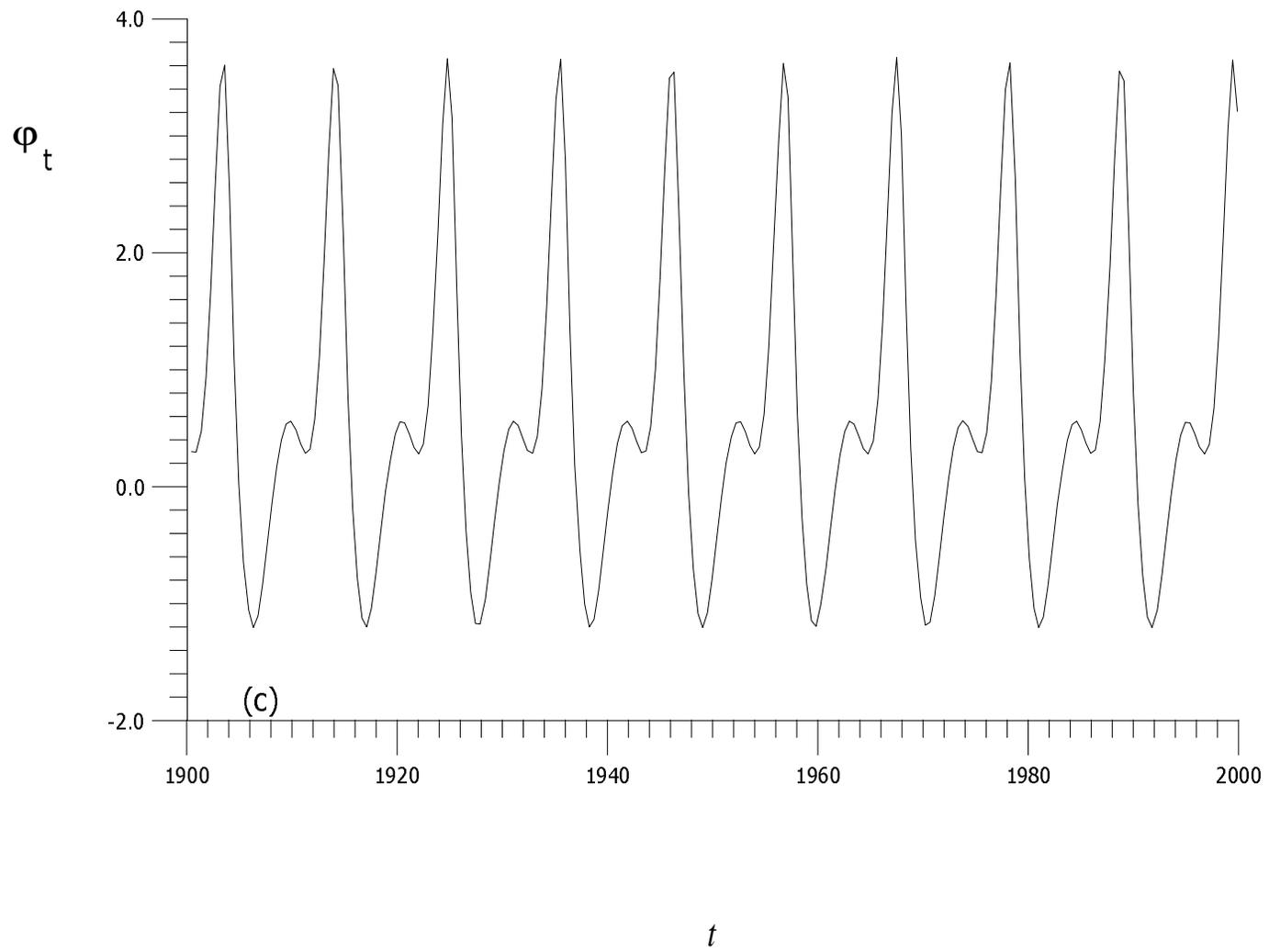
\end{document}